\documentclass[twocolumn,twoside,slac_two]{revtex4}
\usepackage{graphicx}
\usepackage{epstopdf}
\usepackage{fancyhdr}
\newcommand{\met}{\,/\!\!\!\!E_{T}}

\begin{document}

\title{Search for Supersymmetry in $\textrm{p}\overline{\textrm{p}}$ Collisions at $\sqrt{\textrm{s}}$ =1.96 TeV Using the Trilepton Signature of Chargino-Neutralino Production}

%

\author{R. Forrest}
\affiliation{Department of Physics, University of California, Davis, Davis, CA 95616, USA}

\begin{abstract}
The production of chargino-neutralino pairs and their subsequent leptonic decays is one of the most promising supersymmetry (SUSY) signatures at the Tevatron $p\bar{p}$ collider. We present here the most recent results on the search for the three-lepton and missing-transverse-energy SUSY signature using data collected with the CDF II detector. The results are interpreted within the minimal supergravity (mSugra) scenario.
\end{abstract}

\maketitle

\thispagestyle{fancy}


\section{Introduction}
In the search for new phenomena, one well-motivated extension to the Standard Model (SM) is supersymmetry (SUSY). 
SUSY particles (sparticles) contribute to the Higgs mass squared with opposite sign relative to the
contributions of SM particles, and thus protect the weak mass scale, $M_W$, from divergences.  
SUSY is a broken symmetry since the sparticles obviously do not have the
same mass as their SM partners, but the breaking must be `soft' to
allow the divergence canceling to remain effective.  If $R_p$ parity is conserved\footnote{$R_p = (-1)^{3B+L+2S}$, where $B$ is baryon number, $L$ is lepton number, and $S$ is spin.}, the lightest SUSY
particle (LSP) is absolutely stable and provides a viable candidate for cosmological dark matter \cite{susy_primer}. We use as a reference the mSugra model of SUSY breaking.  This model has the virtue of containing only five free parameters to specify.  However, our search is signature-based; we do not modify our selection to follow the details of mSugra.

 One very promising mode for SUSY discovery at hadron colliders is that of
chargino-neutralino associated production with decay into three leptons. Charginos decay into a
single lepton through a slepton $$\tilde{\chi}_1^{\pm}  \rightarrow ~ \tilde{l}^{(*)} ~\nu_l \rightarrow
\tilde{\chi}_1^{0} ~l^{\pm} ~\nu_l $$ and neutralinos similarly decay into two detectable
leptons $$\tilde{\chi}_2^{0}  \rightarrow ~ \tilde{l}^{\pm(*)} ~l^{\mp}
\rightarrow \tilde{\chi}_1^{0} ~l^{\pm} ~l^{\mp} .$$  The decays can also proceed via $W$ and $Z$ bosons.
The detector signature is thus three SM leptons with associated missing energy from the undetected neutrinos and lightest neutralinos, $\tilde{\chi}_1^0$ (LSP), in the event.  Due to its electroweak production, this is one of the few `jet-free' SUSY signatures.

\section{Detector, Data and Analysis Overview}
This analysis is preformed with the CDF II detector at the Tevatron with $p\bar{p}$ collisions at $\sqrt{s} = 1.96$ TeV. The CDF II detector is a mostly cylindrical particle detector composed of cylindrical sub-detectors. From the beam axis outwards there is a silicon strip vertex detector, and a gas filled drift chamber. This tracking system is surrounded by a solenoid providing a 1.4 T magnetic field, followed by electromagnetic and hadronic calorimeters. The outermost detectors are wire chambers used to detect muons that escape the inner detectors.

For this analysis we use two categories of event triggers. The first is the high $P_t$ inclusive lepton trigger, which consist of single lepton objects, the second is the SUSY dilepton trigger witch allows two lower $P_t$ leptons. These data are combined and in the analysis overlapping trigger effects and efficiencies are accounted for. These data were collected up until 1 Jul, 2008, totaling $3.23~\textrm{fb}^{-1}$ for the unprescaled triggers. 

We follow the same analysis strategy and implementation used in the previous CDF II search \cite{rut_note}. From the outset, we define lepton categories and event level trilepton channels. Each lepton and category is exclusive and selected based on expected purity. This channel independence allows easy statistical combination of the final results.

The general procedure is as follows. For each event, we select muons, electrons and tracks of some quality. Each of these objects, except the tracks (T), have tight (t) and loose (l) categories. We then define event level
exclusive trilepton channels composed of combinations of these objects and arrange them sequentially by expected signal sensitivity. There are several virtues of this approach. The largest advantage is that we perform several lepton flavor, channel-specific searches simultaneously, without the need to account for overlapping results.

We define two selection stages to test our background estimations against data. The first stage is the dilepton selection, which consists of the first two objects of the trilepton selection. The second stage is the final trilepton selection, with some event cuts applied. Once we are satisfied with the agreement in the control regions, we apply SUSY specific cuts and look at signal region data to compare against background.

\section{Object Selection, Event Categories and Cuts}
We define both tight and loose lepton categories as well as a track object. All of these objects are central to the detector, meaning that generally $|\eta| < 1.0$ and they are isolated from nearby objects. Tight muons are objects that have tracks, deposit a minimum amount of ionizing energy in the calorimeter system, and are detected in the outer muon systems. Loose muons are similar, but the requirement of the muon system detection is relaxed; they compensate for gaps in the muon detector coverage. Tight electrons are similarly again required to leave a good track, but they are expected to deposit a majority of their energy in the electromagnetic calorimeter. Loose electrons have slightly fewer requirements on the matching between objects in the sub-detectors. We also include one type of track object in the analysis as a possible third object. This greatly increases our sensitivity by allowing detection of leptons that failed selection cuts, as well as single pronged hadronic tau decays.  The track object is a single, isolated track in the tracking chamber. It differs from loose muons in that, it can have an arbitrary amount of energy deposition.

After the object selection, we categorize the trilepton events. We first look for three tight leptons. If the event does not qualify, we look for two tight and one loose lepton. If the event still does not qualify, we look for one tight and two loose leptons. Events that do not make it into the trilepton selection are tested for two tight leptons and a track and finally one tight, one loose and one track object. The complete list is shown in Table ~\ref{table:trilep_select2} along with the $E_t$ (electrons) or $P_t$ (muons) requirements on these objects \footnote{Some selections differ slightly to increase sensitivity or accommodate standard object definitions, see ~\cite{rut_note}.}.

\begin{table}[htbp] 
\begin{center}
\scalebox{0.95}{%
\begin{tabular}{|l|c|c|}
    \hline
Channel & Selection & $E_t$ or $P_t$  \\
\hline
ttt & 3 tight leptons  & 15, 5, 5\\
ttC& 2 tight and 1 loose lepton & 15, 5, 5 \\
tll & 1 tight and 2 loose leptons & 20, 8, 5 \\
\hline
tt$T$ & 2 tight leptons and 1 track & 15, 5, 5 \\
tl$T$ & 1 tight lepton, 1 loose lepton and 1 track & 20, 8, 5 \\
\hline
\end{tabular}}
\end{center}
\caption{Trilepton selection event categories.}
\label{table:trilep_select2}
\end{table}

At this stage we apply additional event level cleaning cuts. We require that every analysis level object (leptons, tracks and jets) be separated from each other by $\Delta R > 0.4 $. Events with a mismeasured jet can have false $\met$. We remove events with $\met$ and any jet separated by less than $\Delta \phi < 0.35$. We also make invariant mass cuts at this stage. The highest opposite signed object pair invariant mass is required to be above $20\ \textrm{GeV}/c^2$ and the second highest oppositely-charged object pair is required to be above $13\	 \textrm{GeV}/c^2$. This cut helps eliminate heavy flavor backgrounds.

Additional backgrounds due to mismeasurement are removed by cutting events that have $\met$ and leptons aligned,  requiring $\Delta \phi > 0.17$ for each of the leading two leptons. 

To further clean up events, we require the third lepton in trilepton events to be isolated.  We also require that there not be more than three leptons or tracks in the event above 10 GeV and that the three objects' charges sum to $\pm 1$.

\subsection{Backgrounds}
The standard model background estimation for the analysis differs slightly between the lepton+track channels and the trilepton channels. Generally, Monte Carlo is used to estimate the backgrounds, but isolated track, fake lepton and gamma conversion rates are determined from data.
\subsection{Trilepton Backgrounds}
Backgrounds are treated differently based on the underlying process. Those that give three real leptons (WZ , ZZ, $t\bar{t}$) are estimated with Monte Carlo by simply taking them through the analysis.

 The remaining background processes have two real leptons (Z, WW) and require a third object from elsewhere in the event. This can happen, for example, with FSR photon conversion where a photon radiated off a charged particle hits matter in the detector and converts to an $e\bar{e}$ pair. For these processes, we estimate this 2 lepton plus conversion rate from Monte Carlo.
 
The final contribution to the trilepton background is from objects in the underlying event faking a third lepton in an event that has two genuine leptons. This fake contribution is estimated in the trilepton channels by selecting two well identified leptons and a third fakeable object from data events. Fake rates have been measured for jets faking electrons and for tracks faking muons of both tight and loose quality. These jets and tracks are the fakeable objects selected. The event is then carried through the analysis and weighted by the appropriate fake rate.

\subsection{Dilepton + Track Backgrounds}
For channels with tracks, backgrounds are handled slightly differently. Background processes that give three real leptons (WZ , ZZ, $t\bar{t}$) are still estimated from Monte Carlo as previously described. 

As for fakes in dilepton + track channels, we account for fake leptons, and separately estimate the rate of isolated tracks in dilepton backgrounds. 
 
For fake leptons, we use a method similar to the trilepton method but calculated from data by selecting lepton + track events containing a fakeable object. As was done with trilepton fakes, we carry the event through the analysis, and apply the appropriate fake rate to the event. 

 The remaining background in the dilepton + track channels is that of dilepton events with an isolated track from the underlying event. We measure the rate of extraneous isolated tracks from data, and apply this rate to dilepton Monte Carlo. This procedure gives very good agreement in our dilepton + track control regions.
 
 \section{Control Regions}
\begin{figure}
 \begin{center}
   \includegraphics[width=3.5in]{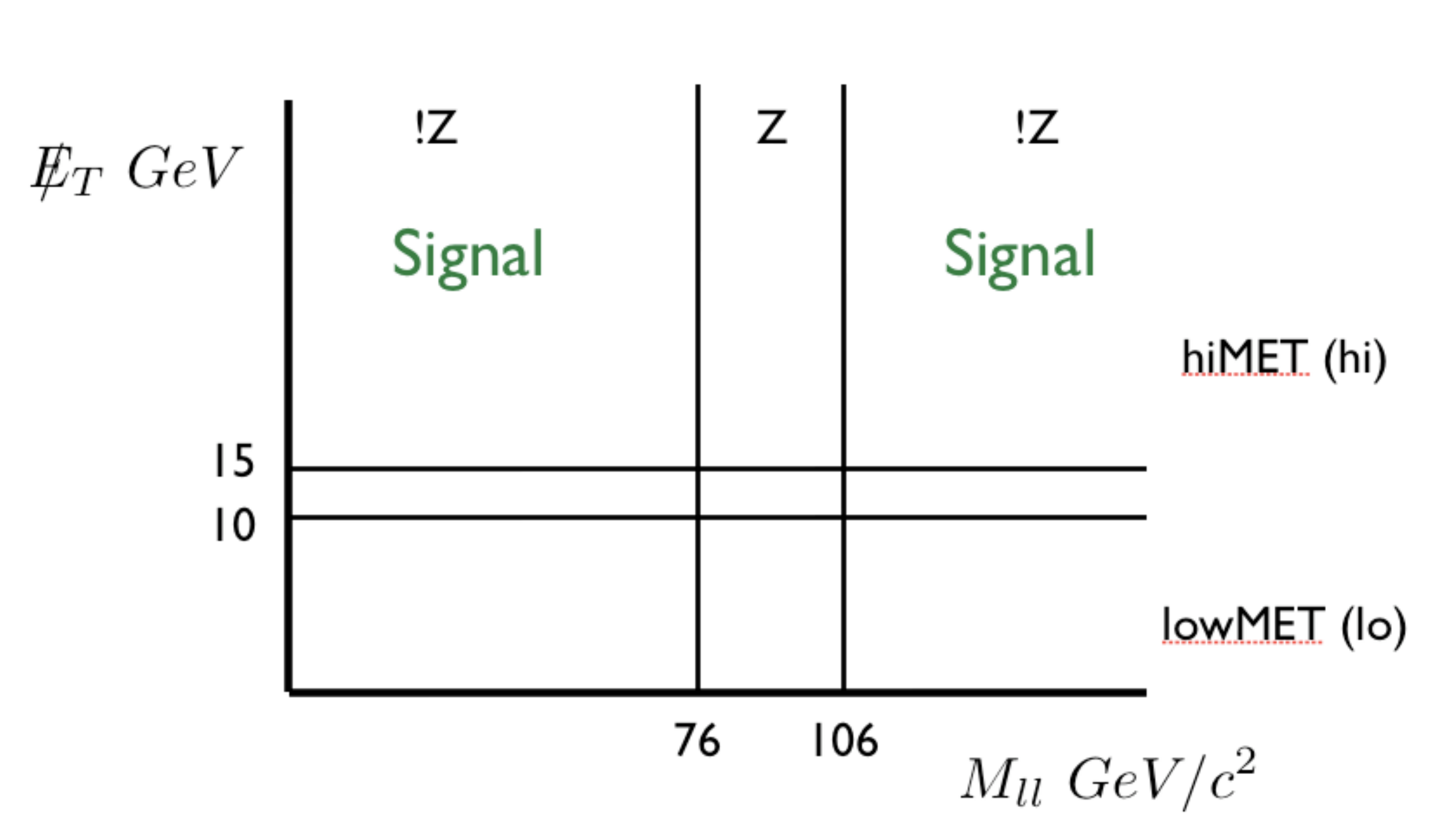}
 \end{center}
     \caption{Control regions and codes used to refer to the control regions.}
     \label{fig:control_regions}
\end{figure}

 We inspect both our dilepton selection and our trilepton selection for agreement against predictions. The control region parameter space is $\met$ vs. Invariant mass, and for easy reference is coded according to Figure ~\ref{fig:control_regions}.

\begin{figure}
 \begin{center}
   \includegraphics[width=3.5in]{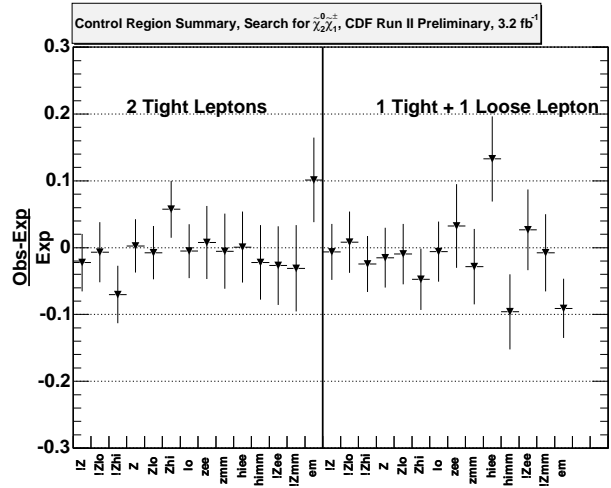}
 \end{center}
     \caption{Summary of dilepton control regions. (Observed - Expected) / Expected number of events for each control region.}
     \label{fig:dilep_summary}
\end{figure}

\begin{figure}
 \begin{center}
   \includegraphics[width=3.0in]{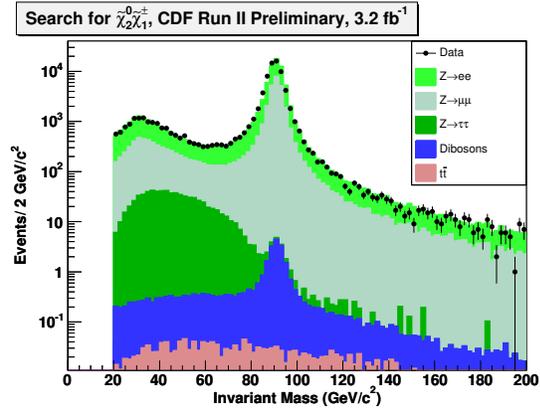} 
 \end{center}
     \caption{Invariant mass of the first two tight leptons in events with low $\met$.}
     \label{fig:dilep_plots}
\end{figure}

We select the first two leptons in the event and check agreement against backgrounds. See Figure ~\ref{fig:dilep_summary} for a complete listing of all the dilepton control regions. A dilepton kinematic plot is displayed and described in Figure ~\ref{fig:dilep_plots}.

\begin{figure}
 \begin{center}
   \includegraphics[width=3.5in]{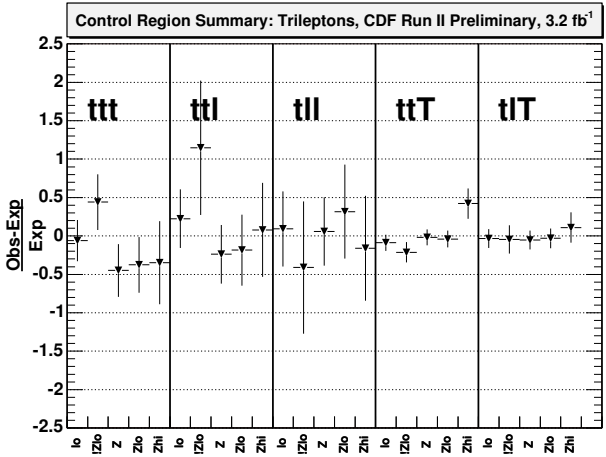}
 \end{center}
     \caption{Trilepton Control Region Summary. (Observed - Expected)/Expected number of events.}
     \label{fig:trilep_cr_summary}
\end{figure}

After we are satisfied with the dilepton control region agreement, the trilepton selection is applied to an event. We check trilepton plots and tables to ensure good agreement between background and predictions. The total trilepton background and prediction comparison is shown in Figure \ref{fig:trilep_cr_summary}. 

\begin{figure}
 \begin{center}
   \includegraphics[width=3.0in]{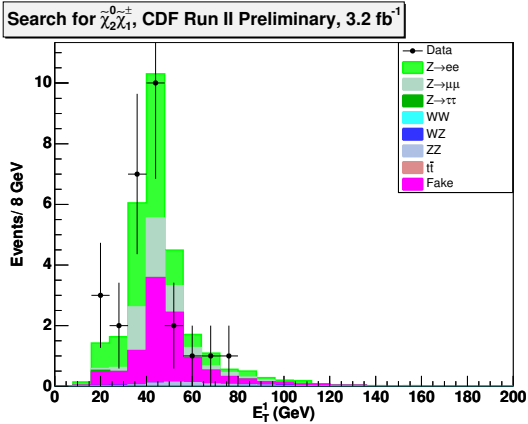} \\
 \end{center}
     \caption{Energy of leading lepton in events with low $\met$ in the ttt channel. }
       \label{fig:trilep_plots}
\end{figure}

We again look at distributions comparing data and predictions in control regions.  Trilepton control region plots are shown in Figure ~\ref{fig:trilep_plots}.

\section{Results and Limits}

For an mSugra reference point we use  $\textrm{M}_0 = 60, \textrm{M}_{1/2} = 190, \textrm{tan} \beta = 3, \textrm{A}_0 = 0$; the results of background and expected signal are shown in Table \ref{table:exp_sig_sys}. After looking at the signal region in the data, we see a total of seven signal events on an expected background of $10.84 \pm 1.34$ events.

\begin{table}[htbp] 
\begin{center}
\scalebox{0.8}{%
\begin{tabular}{|c|c|c|c|}
\hline
\multicolumn{4}{|l|}{CDF II Preliminary, 3.2 $\textrm{fb}^{-1}$} \\
\hline
Channel& Total Background $\pm$ (stat) $\pm$ (sys)& Signal Point $\pm$ (stat) $\pm$ (sys)&Observed\\
\hline
ttt &0.83 $\pm$ 0.14 $\pm$ 0.11  & 3.64  $\pm$ 0.22 $\pm$ 0.49&1\\ 
ttC & 0.39 $\pm$ 0.07 $\pm$ 0.04 & 2.62 $\pm$ 0.18 $\pm$ 0.35  &0\\
tll &0.25  $\pm$ 0.08 $\pm$ 0.03 & 1.12 $\pm$ 0.12 $\pm$ 0.15  &0\\
ttT &5.85  $\pm$ 0.57 $\pm$ 1.11& 7.15 $\pm$ 0.31$\pm$ 0.91  &4\\
tlT &3.53  $\pm$ 0.52 $\pm$ 0.5 &4.06 $\pm$ 0.23  $\pm$ 0.53 &2\\
\hline
\multicolumn{4}{l}{mSugra Signal point: $\textrm{M}_0 = 60, \textrm{M}_{1/2} = 190, \textrm{tan} \beta = 3, \textrm{A}_0 = 0$} \\

\end{tabular}}
\end{center}
\caption{Expected background and signal, errors are statistical and full systematic.}
\label{table:exp_sig_sys}
\end{table}

\begin{figure}
 \begin{center}
   \includegraphics[width=3.5in]{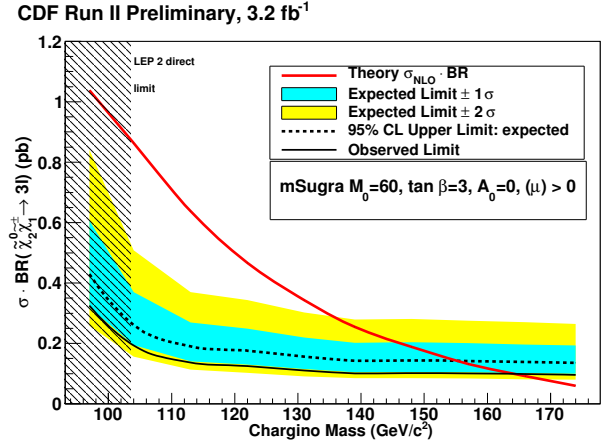}
 \end{center}
     \caption{Expected and observed limit for the mSugra model $M_0 = 60, tan \beta = 3, A_0 = 0, (\mu) > 0$. In red is the theoretical $\sigma \times BR$ and in black is our expected limit with one and two $\sigma$ errors. We expect to set a limit of about 156 GeV/$c^2$, and observe a limit of 164  GeV/$c^2$.}
     \label{fig:exp_sig_1D}
\end{figure}

To extract a 1-D 95\% confidence level limit, we set $M_0 = 60$ and vary $M_{1/2}$ which has the direct effect of varying the chargino mass. For each point we scan, we get the expected limit based on the acceptance of our analysis to the signal at that point. If we plot this against the theoretical $\sigma \times BR$ of the signal mSugra point as a function of chargino mass, we expect to exclude regions where our analysis's $\sigma \times BR$ is less than the theoretical value. Our expected limit is about 156 GeV/$c^2$  Figure \ref{fig:exp_sig_1D}, while we observe a limit of 164  GeV/$c^2$.

To explore a broader parameter space it is useful to scan both $M_0$ and $M_{1/2}$ simultaneously. We calculate NLO cross section of the process as a function of  $M_0$ and $M_{1/2}$.  We then calculate branching ratio to three leptons in this same range. This gives us a plot of $\sigma  \times BR$. We generate signal Monte Carlo to test the expected and observed sensitivity at many points in $M_0$ and $M_{1/2}$ space.

\begin{figure}
 \begin{center}
   \includegraphics[width=3.5in]{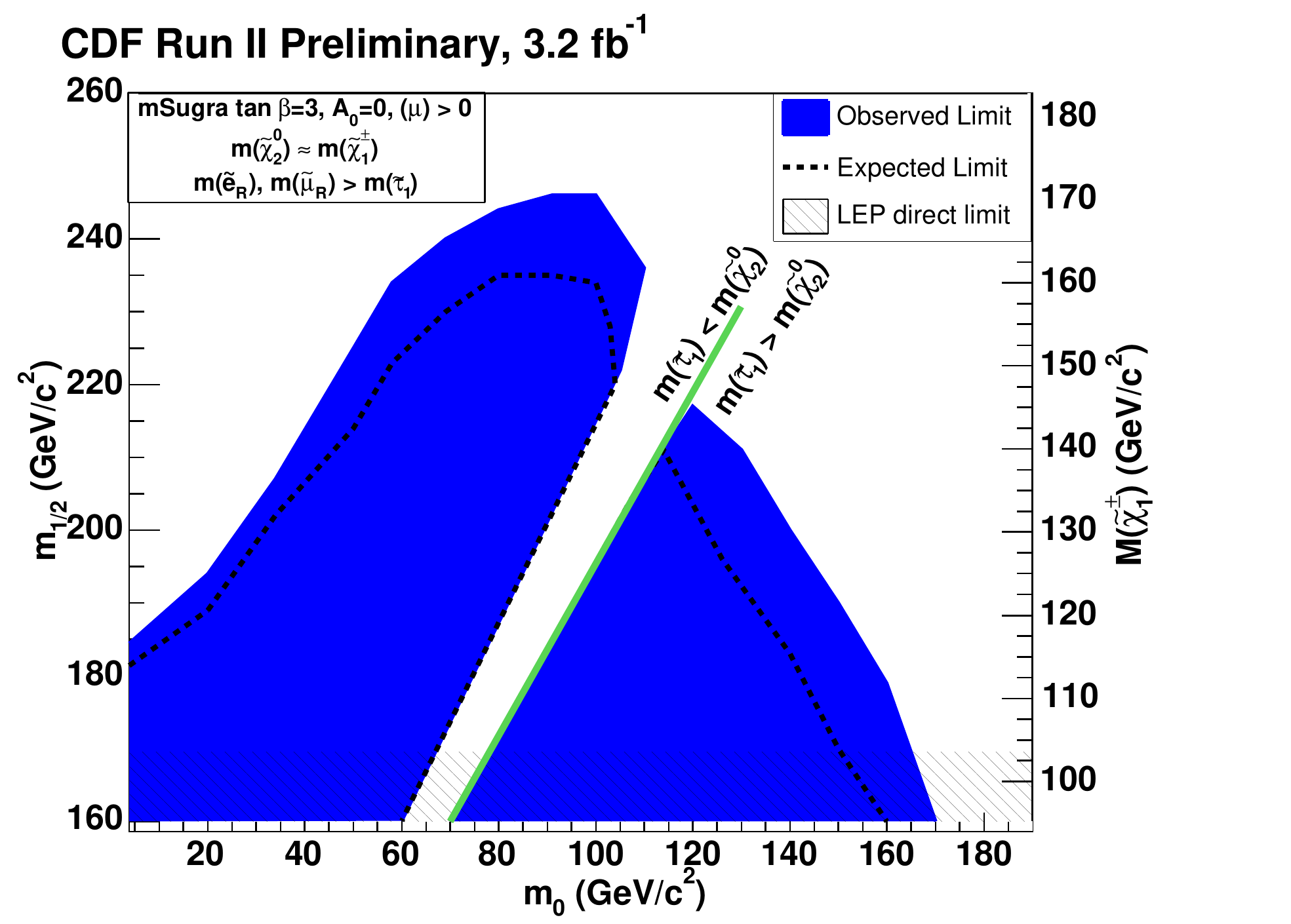}
 \end{center}
     \caption{Expected and observed limit contours for the mSugra model  $tan \beta = 3, A_0 = 0, (\mu) > 0$ in $M_{1/2}$ vs $M_{0}$ space.}
     \label{fig:exp_sig_2D}
\end{figure}

We calculate (Expected - Theory $\sigma \times$  BR) /(Theory $\sigma \times$ BR) for both the expected and observed limits. The final exclusion contains both of these contours which can be seen in Figure \ref{fig:exp_sig_2D}.

 Our observed 1-D limit excludes chargino masses of less than $164\ \textrm{GeV}/c^2$, an improvement over the expectation due to the deficit of data events in the lepton + track channels. 

\bigskip 

\end{document}